\documentclass[aps,a4paper,twocolumn,amsmath,showpacs,prb]{revtex4}

\usepackage{graphicx}

\begin{document}
\title{Time evolution of dynamic propensity in a model glass former.\\ The interplay between structure and dynamics.}
\author{J. A. \surname{Rodriguez Fris}}
 \email{rodriguezfris@plapiqui.edu.ar}
\author{L. M. Alarc\'on}
\author{G. A. Appignanesi}
 \affiliation{
Fisicoqu\'{\i}mica, Departamento de Qu\'{\i}mica, Universidad Nacional del Sur, Av. Alem 1253, 8000 Bah\'{\i}a Blanca, Argentina.\\}
\date{\today}


\begin{abstract}

By means of the isoconfigurational method we calculate the change in the propensity for motion that the structure of a glass-forming system experiences during its relaxation dynamics. The relaxation of such a system has been demonstrated to evolve by means of rapid crossings between metabasins of its potential energy surface (a metabasin being a group of mutually similar, closely related structures which differ markedly from other metabasins), as collectively relaxing units (d-clusters) take place. We now show that the spatial distribution of propensity in the system does not change significantly until one of these d-clusters takes place. However, the occurrence of a d-cluster clearly de-correlates the propensity of the particles, thus ending up with the dynamical influence of the structural features proper of the local metabasin. We also show an important match between particles that participate in d-clusters and that which show high changes in their propensity.

\end{abstract}


\maketitle

\section{Introduction}

The determination of the physical basis of the emergence of glassy relaxation when a liquid is supercooled under its melting point represents a subject of intense research \cite{angell_91,ediger_00,debenedetti_01,ediger_96,gotze_99,proceedings_02}. Striking aspects of such behavior are the fact that the dynamic observables change dramatically with the supercooling (while static quantities show at most very mild differences) and that the dynamics varies by orders of magnitude from one region of the system to another \cite{angell_91,ediger_00,debenedetti_01,ediger_96,gotze_99,proceedings_02}. The dynamics of these systems slows down very fast in this regime (as temperature is reduced) and the relaxation has been believed to proceed by means of cooperatively relaxing regions whose time scales and sizes grow considerably as the temperature is decreased \cite{angell_91,ediger_00,debenedetti_01,ediger_96,gotze_99,proceedings_02,adam_65}. This heterogeneous scenario has been corroborated by the discovery, both experimentally and theoretically, of the existence of dynamical heterogeneities \cite{schmidt_91,donati_98,richert_02,weeks_00,kegel_00,kob_97,butler_91,cicerone_95}. Some studies identified cooperative motions of a small number of particles (a few percent) which move collectively, often in a string-like fashion, by a distance close to the particle diameter \cite{kegel_00,weeks_00,richert_02,donati_98}. More recently, collective motions of a significant fraction of the particles of different regions of these systems which form relatively compact clusters have been found \cite{appignanesi_06}. These very rapid and sporadic events, which were termed ``democratic motions'', drive the system from one metabasin (MB) of its potential energy surface, a group of similar closely related structures \cite{appignanesi_06,debenedetti_01,vogel_04}, to another while the structural relaxation (the so-called $\alpha$ relaxation) is performed by a small number of such events.
These relatively compact clusters (``democratic'' clusters or d-clusters \cite{appignanesi_06}) have been identified in different glassy systems like a binary Lennard-Jones system and supercooled water \cite{agua_07} and represent natural candidates for the cooperatively relaxing regions proposed long ago by Adam and Gibbs \cite{adam_65}. A recent inhomogeneous mode-coupling theory of dynamical heterogeneity has related them to the (fractal) geometrical structures carrying the dynamical correlations at timescales commensurable with that of the $\alpha$ relaxation \cite{biroli_06}. Additionally, a recent experimental and computational work in a glassy polymer provided experimental support to the MB-MB transitions and d-clusters \cite{vallee_07}.

Even when it would be intuitively expected that the dynamical heterogeneities are related to structural heterogeneities in the sample, the determination of the existence of a causal link between structure and dynamics remains as another main open problem in the field \cite{ediger_00,widmer-cooper_04,widmer-cooper_07}. A recent beautiful idea holds the promise to shed some light in this regard \cite{widmer-cooper_07,widmer-cooper_04}. By means of the isoconfigurational ensemble (IC) it has been determined that the propensity of the particles (their tendency to be mobile) is in fact determined by the initial configuration of the sample and that the particles with the higher propensity are not homogeneously distributed within the sample but arranged in relatively compact clusters \cite{widmer-cooper_07,widmer-cooper_04}. The time extent of the influence of the local structure on the dynamics has also been shown to be on the order of the MB residence time, a timescale shorter than the $\alpha-$relaxation time ($\tau_\alpha$) \cite{PRL_2}. In a previous work, it has been shown for the binary Lennard-Jones system that the high propensity regions of a given initial configuration represent unblocked regions wherein d-clusters occur in the subsequent dynamics (for any given isoconfigurational realization or trajectory initiated in such configuration) \cite{frechero_06}.

The aim of the present work is to complete this picture by revealing the fact that not only the local structure constraints the resulting dynamics but to elucidate the role of the d-clusters in reformulating the spatial variation of propensity for motion, thus making evident the mutual transference of constraints between structure and dynamics. To this end, and by means of molecular dynamics (MD) simulations of the binary Lennard-Jones model we shall generate many ICs at different times over a given trajectory (thus determining the spatial distribution of propensity on configurations at different times of the trajectory) and we shall show that at the times when the d-clusters occur the system shows the greater changes in propensity, thus producing a clear propensity de-correlation.

\section{Model system and methods}

\subsection{The binary Lennard-Jones system}

We performed a series of MD NVE simulations for a widely used model of fragile glass former: the binary Lennard-Jones system consisting of a 3D mixture of 80 \% $A$ and 20 \% $B$ particles, the size of the $A$ particles being 47 \% larger than the $B$ ones \cite{donati_98,appignanesi_06,footnote1}. We shall show results from systems at temperature $T=0.5$, density of 1.2 and $N=150$ particles \cite{appignanesi_06}. This system size avoids the interference of results from many different subsystems (metabasins) while being free of finite size effects, as shown in Ref.~\onlinecite{appignanesi_06}. We have also simulated a large system ($N=8000$ particles) which yielded an equivalent dynamical behavior (however, in order to look for MBs and d-clusters, one has to study the behavior of subsystems of $N=150$ and thus, the large system should be decomposed in many small subsystems wherein to apply the concept of MB \cite{doliwa_03,vogel_04}). Given this fact, in the rest of this work we shall refer to a system of $N=150$. At low temperatures (close and above the critical temperature predicted by the mode-coupling theory of the glass transition, $T_{\rm c}=0.435$) this system presents dynamical heterogeneities \cite{donati_98,appignanesi_06}: a small number of particles move cooperatively a distance that is comparable to the inter-particle distance. These ``fast moving'' (or ``mobile'') particles are not homogeneously distributed throughout the sample but are arranged in clusters usually made of string-like groups of particles \cite{donati_98,appignanesi_06}. The dynamics is most heterogeneous at time $t^*$ defined by the maximum in the non-gaussian parameter $\alpha_2(t)$, $\alpha_2(t)= \frac{3 \langle r^4(t) \rangle}{5 \langle r^2(t) \rangle^2} - 1$, which measures the deviation of the self part of the van Hove function (the probability at a given time of finding a particle at distance $r$ from its initial position) from a brownian behavior \cite{donati_98}. This quantity is located at the end of the $\beta$ - beginning of the $\alpha$ relaxation (the crossover from the caging to the diffusive regime in the mean-squared displacement $\langle r^2(t) \rangle$ plot) and constitutes the characteristic time for dynamical heterogeneities (in this case $t^*=400$ for $T=0.50$ \cite{appignanesi_06}). Additionally, $t^*$ depends strongly on temperature and grows quickly as we move towards $T_{\rm c}$ \cite{kob_95}. However, not all the mobile particles within a $t^*$ time span contribute decisively to the $\alpha$ relaxation, as we have recently demonstrated \cite{appignanesi_06}. Instead, the $\alpha$ relaxation is driven by a series of a few MB-MB transitions which are triggered by the occurrence of large compact clusters of medium-range-mobility particles called democratic particles \cite{appignanesi_06}. Additionally, the mean-residence time in an MB has been estimated to be close to $t^*$ \cite{appignanesi_06}.

\subsection{The distance-matrix method, metabasins and d-clusters}
\label{IIb}

We now describe briefly the distance-matrix method to study MB dynamics (see Refs.~\onlinecite{appignanesi_06,PRL_2} for details): We perform an MD simulation and record equally-spaced configurations (for example 101 configurations, as in Refs.~\onlinecite{appignanesi_06,PRL_2}) for an $\alpha$-relaxation total run time ($\tau_\alpha = 4000$, thus consecutively recorded configurations are separated in time by $\phi=10$ \% $t^*=40$) and build the following distance matrix \cite{ohmine_95,appignanesi_06}, $\Delta^2(t',t'') = \frac{1}{N}\sum_{i=1}^N |{\bf r}_i(t')-{\bf r}_i(t'')|^2$, where ${\bf r}_i(t)$ is the position of particle $i$ at time $t$ (since $\Delta^2(t',t'') = \Delta^2(t'',t')$ this is a triangular matrix). Thus $\Delta^2(t',t'')$ gives the system-averaged-squared displacement of a particle in the time interval that starts at $t'$ and ends at $t''$. In other words, this distance matrix contains the averaged-squared distances between each recorded configuration and all the other ones. For this study (as all studies dealing with MBs \cite{appignanesi_06,ohmine_95,vogel_04}), we must investigate small systems, since for large systems the results originated from different subsystems would obscure the conclusions \cite{appignanesi_06,ohmine_95,vogel_04}. Thus, we used 150 particles. However, we also found the same qualitative results for small subsystems immersed in a big one, thus ruling out the possibility for finite size effects (we repeated the study for subsystems of 150 particles within a large system of 8000 particles, that is, we focused only on a portion of the large system).

\begin{figure}[h!]
\includegraphics[width=1\linewidth]{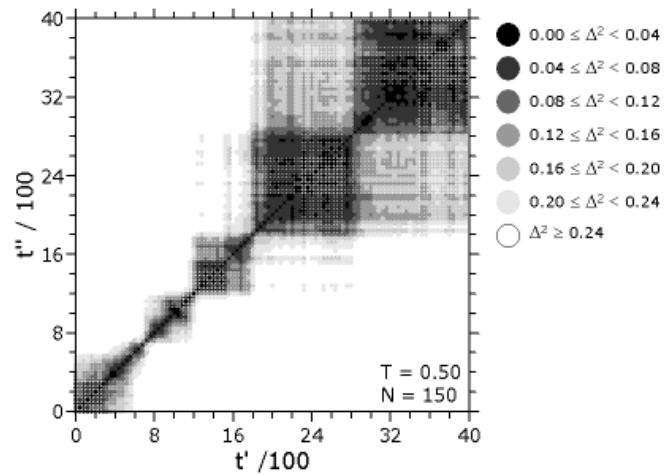}
\caption{Contour plot of the distance matrix $\Delta^2(t',t'')$ for $T=0.5$. The gray levels correspond to the values that are given to the right. From Ref. \onlinecite{appignanesi_06}.}
\label{fig1}
\end{figure}

Figure \ref{fig1} shows a typical behavior for trajectories with $T=0.5$.
The gray level of the squares in $\Delta^2(t',t'')$ depicts the distance between the corresponding configurations, the darker the shading indicating the lower the distance between them. Were the dynamics homogeneous in time, we would expect a distance matrix with a dark main diagonal and a continuous fading as we move away from such diagonal. However, from the island structure of this matrix a clear MB structure of the landscape is evident. That is, islands are made up of closely related configurations (low $\Delta^2$) which are separated from the configurations of other islands by large distances. In other words, the dynamics of the system is inhomogeneous in time. We can estimate the typical residence time in the MBs for this $T$ (from island sizes) as qualitatively on the order of $t^*$. Given the small system size we expect this to be a good estimate (however, this timescale clearly depends on system size, since for a large system different subsystems would be undergoing MB-MB-transition events at different times). Thus, MB-MB transitions (the crossings from one island to another and which last 10 \% to 20 \% of $t^*$) are fast events compared to the times for the exploration of the MBs. The study of MB-MB-transition events has been done previously \cite{appignanesi_06}, revealing the decisive role of large compact d-clusters of medium-range-mobility particles. These clusters are responsible for the $\alpha$ relaxation (completed after 5 - 10 such events) and represent potential candidates for the cooperatively relaxing regions of Adam \& Gibbs \cite{appignanesi_06}. The compact nature of these events relevant to the $\alpha$ relaxation has been shown to be in accord with the geometrical structure of the dynamics correlations at large timescales on the order of $\tau_\alpha$ within an inhomogeneous mode-coupling theory (at variance from the less dense structures, compatible with string-like motions, expected at the shorter timescales of $\beta$-relaxation time) \cite{biroli_06}.

The democratic particles that comprise the d-clusters that trigger MB-MB transitions were defined as that whose mobility was greater than $r_{th}=0.3$ within the time interval $\phi=40$, and its fraction is represented by the function $m(t,\phi)$ \cite{appignanesi_06}. Thus, for the system size under study we found that on the order of 40 - 60 particles were involved in a d-cluster \cite{appignanesi_06}.

\begin{figure}[h!]
\includegraphics[width=1\linewidth]{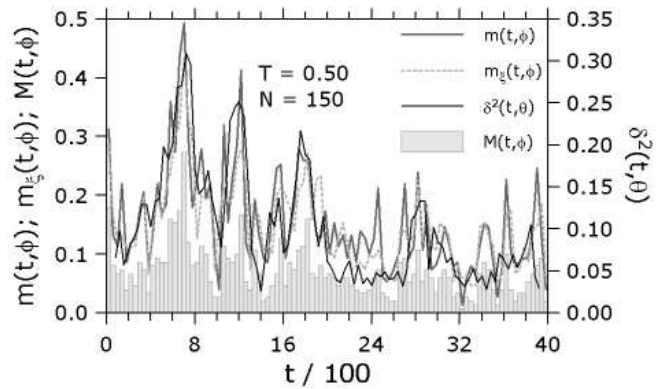}
\caption{
Solid-black line (right ordinate): Averaged-squared displacement $\delta^2(t,\theta)$ for the trajectory given in Fig.~\ref{fig1}.
Solid-dark-gray line (left ordinate): The function $m(t,\phi)$ which gives the fraction of democratic particles, i.e., particles that moved more than the threshold value $r_{th}= 0.3$ in the time interval $[t,t+\phi]$.
Dashed-light-gray line (left ordinate): The function $m_\xi(t,\phi)$ which gives the fraction of particles with a great propensity-change, i.e., particles that changed their propensity more than the threshold value $\xi = 0.21$ in the time interval $[t,t+\phi]$.
Vertical bars (left ordinate): The function $M(t,\phi)$ with gives the fraction of particles that are both democratic and have $\xi > 0.21$. The values of $\theta$ and $\phi$ are respectively 160 and 40.}
\label{fig2}
\end{figure}

In Fig.~\ref{fig2} we show for the same trajectory and total run time interval the function $\delta^2(t,\theta)$, the system-averaged-squared displacement of the particles within a time interval $\theta$ (black-solid curve). This function is defined as:
\begin{eqnarray*}
\delta^2(t,\theta)  & = & \Delta^2(t-\theta/2,t+\theta/2) \\
 & = & 
\frac{1}{N}\sum_{i=1}^N |{\bf r}_i(t-\theta/2)-{\bf r}_i(t+\theta/2)|^2 \,.
\end{eqnarray*}
\noindent
Thus $\delta^2(t,\theta)$ is $\Delta^2(t',t'')$ measured along the diagonal $t''=t'+\theta$ and hence the average of this quantity over different start times $t$ gives the usual mean-squared displacement for time lag $\theta$. For this plot we have chosen $\theta=160$, a value that is significantly smaller than the $\alpha-$relaxation time
($\tau_\alpha=4000$) but still sufficiently larger than the time of the microscopic vibrations (=$O(1)$). A comparison of this $\delta^2(t,\theta)$ with Fig.~\ref{fig1} shows that $\delta^2(t,\theta)$ is showing pronounced peaks exactly then when the system leaves an MB. Thus we see that changing the MB is indeed associated with a rapid motion as measured in $\delta^2(t,\theta)$.
In Fig.~\ref{fig2}, we have included the fraction of democratic particles $m(t,\phi)$ \cite{appignanesi_06} as a function of time (dark-gray-solid line). The comparison with $\delta^2(t,\theta)$ shows that the fraction of these particles is indeed large whenever the $\delta^2(t,\theta)$ increases rapidly. This fraction is on the order of 30 \% to 40 \% of all the particles \cite{appignanesi_06} and, thus, significantly larger than one would expect from $4\pi r^2 G_{\rm s}(r,\phi)$ if one integrates this distribution from $r_{th}$ to infinity (which gives $\sim\!0.16$).

\subsection{The isoconfigurational method. The role of the local structure: Dynamic propensity}
\label{IIc}

To calculate the propensity for motion we use the isoconfigurational (IC) method introduced in Ref.~\onlinecite{widmer-cooper_04}. In it one performs a series of equal length MD runs (trajectories) from the same initial configuration, that is, always the same structure (the same particle positions) but each trajectory with different initial particle momenta chosen at random from the appropriate Maxwell-Boltzmann velocity distribution (that is, one builds an IC ensemble). For any given time, each run or trajectory presents certain mobile particles or dynamical heterogeneities. However, the mobile particles and corresponding clusters of mobile particles differ from run to run since the mobility of the particles in a single run is not determined by the initial configuration \cite{widmer-cooper_04}.
Propensity of a particle for motion in the configuration at time $t$ (its tendency to be mobile at the instantaneous time $t$ in the trajectory given in Fig.~\ref{fig1}) for a fixed time interval of length $\phi$ is defined as $\langle \Delta {\bf r}_i^2(t) \rangle_{\rm IC} = P_i(t)$, where $\langle \cdots \rangle_{\rm IC}$ indicates an average over the IC generated at time $t$ and $\Delta {\bf r}_i^2(t) = |{\bf r'}_{\!i}(t,\phi) - {\bf r}_i(t)|^2$.
Here, ${\bf r}_i(t)$ is the position of particle $i$ in the configuration at time $t$ of the trajectory given in Fig.~\ref{fig1} and where one of the ICs is started, and ${\bf r'}_{\!i}(t,\phi)$ is the position of the same particle at the end of one of such IC trajectories of length $\phi$. This definition of propensity is a generalization of the original definition \cite{widmer-cooper_04} since in this work we are interested in studing how propensity evolves with time during a given MD trajectory and to relate such propensity changes to the dynamical events that occur in such trajectory.
At low temperatures propensities for any given configuration at time $t$ are not uniform throughout the sample and high propensity particles are confined to certain (relatively compact) regions \cite{widmer-cooper_04,PRL_2}. Thus, while particle mobility is not reproducible from run to run, the spatial variation in the propensity is completely determined by the initial configuration, reflecting the influence of structure on dynamics \cite{widmer-cooper_04,PRL_2}. While the mean value of the propensity [$\langle P(t) \rangle = N^{-1}\sum_{i=1}^N P_i(t)$] depends on the length of the time interval in which it is calculated ($\phi$), the spatial variation of propensity does not depend on it for times not too small (down to 10 \% $t^*$ or even less) \cite{ARF07,widmer-cooper_07}. In other words, if one calculates the propensity at $\phi=t^*$ or higher (of course that for very long times larger than $\tau_\alpha$ when the system becomes diffusive, the propensity distribution gets uniform) or if one does it for a timescale of 10 \% $t^*$, the particles with high and low propensity are the same in any case. In this work we chose a time $\phi=40=10$ \% $t^*$ to calculate propensity since at this short time the IC trajectories do not have time to abandon the local MB (perform a d-cluster) and we can safely say that we are sampling the short-time vibrations (without interference of the dynamics at larger times). A propensity much greater than the mean value, a great tendency to be mobile, is thus a clear indication that the particle is not ``comfortable'' in its actual position, that is, it is structurally unjammed.

\section{Results}

\begin{figure}[h!]
\includegraphics[width=1\linewidth]{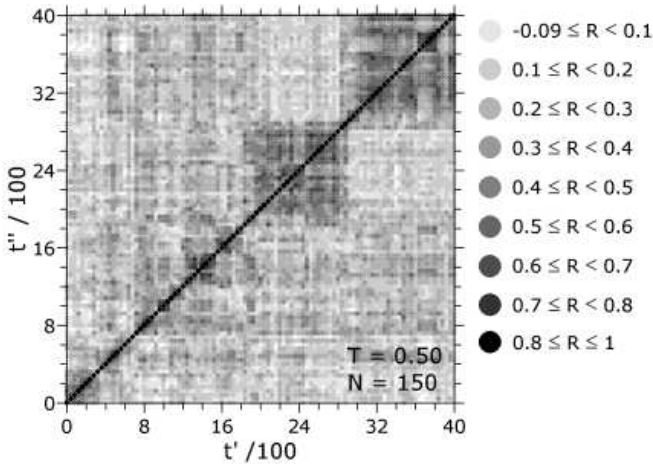}
\caption{Cross-correlation function $R(t',t'')$ between propensities calculated over ICs generated from configurations at times $t'$ and $t''$ that belong to the trajectory given in Fig.~\ref{fig1}. The propensities are calculated for a $\phi=40$ time interval. The gray levels correspond to the values that are given to the right.}
\label{fig3}
\end{figure}

With the tools already described (subsection \ref{IIc}) and in order to elucidate the influence that the dynamical events (subsection \ref{IIb}) have on the local structure of the system, we generated a single dynamical trajectory (Fig.~\ref{fig1}) and started many ICs {\it over} it. That is, we stored configurations from an MD run each $\phi$ time units (a total of 101 configurations, thus the run was for a total length of $\tau_\alpha=4000$) and generated 101 different ICs of 500 trajectories from each of them. Thus, we determined the propensities for each of the 101 equally spaced configurations over the given MD run (we mention that we have obtained similar values of propensity using 200 trajectories for each IC instead of 500). The way the spatial distribution of propensity (the values of the propensity for each of the different particles) changes from one configuration to another, gives an idea of the time evolution of the local structural constraints. To quantify how similar or different is the propensity of the different particles in configurations at times $t'$ and $t''$, we calculate the following cross-correlation function $R=\sum_{i=1}^N R_i$, where:
\[
R_i =
\frac{[X_i - \langle X \rangle]}{\left\{ \sum_{l=1}^k[X_l - \langle X \rangle]^2 \right\} ^{1/2}}
\cdot
\frac{[Y_i - \langle Y \rangle]}{\left\{ \sum_{l=1}^k[Y_l - \langle Y \rangle]^2 \right\} ^{1/2}} \, .
\]
In it, $X_i$ and $Y_i$ are respectively $P_i(t')$ and $P_i(t'')$. $P_i(t)$ is the propensity of particle $i$ calculated over the IC generated from the configuration at time $t$, in the trajectory seen in Fig.~\ref{fig1}. Besides $\langle X \rangle$ and $\langle Y \rangle$ are respectively $\langle P(t') \rangle$ and $\langle P(t'') \rangle$.
Then, with all pairs of times $t'$ and $t''$ we generate the squared matrix $R(t',t'')$. Figure \ref{fig3} depicts $R(t',t'')$, which shows a structure which matches that of the MBs of Fig.~\ref{fig1}. We can see that large values of $R$ (dark regions) occur at times when Fig.~\ref{fig1} shows a metabasin structure. This means that the propensity of the particles within an MB are positively correlated, that is to say, the propensity of each particle is similar in all the corresponding ICs. Since $R$ is generally small for ICs not belonging to the same MB, the propensity of the particles at such times do not present a neat correlation. Moreover, the borders of the squared islands of Figs.~\ref{fig1} and \ref{fig3} show certain small negative values, thus indicating a slight anticorrelation in propensity once the system abandons the local MB (high-propensity particles change to low propensity, or intermediate propensity, and vice versa). This fact means that MB-MB transitions and d-clusters reformulate the propensity pattern, by rearranging the regions of high, medium and low propensity. This is also consistent with a rapid loss of the memory of the local structure (the structural constraints of the present MB) once the system escapes from the MB. We also note that if we calculate propensities at larger times (say, for example at $\phi=t^*$ instead of $\phi=40$), the results are consistent, and even more conspicuous.

We recall that the function $R(t',t'')$ of Fig. \ref{fig3} is built upon ``structural information'' while $\Delta^2(t',t'')$ of Fig. \ref{fig1} relies on dynamical facts. $R(t',t'')$ is based on the propensity for motion of the particles calculated at very short-time intervals $\phi=40$ (at times $t'$ and $t''$) and thus reflects the cross-correlations between the local structural constraints of the configurations at the corresponding times.
 
\begin{figure}[h!]
\includegraphics[width=.8\linewidth]{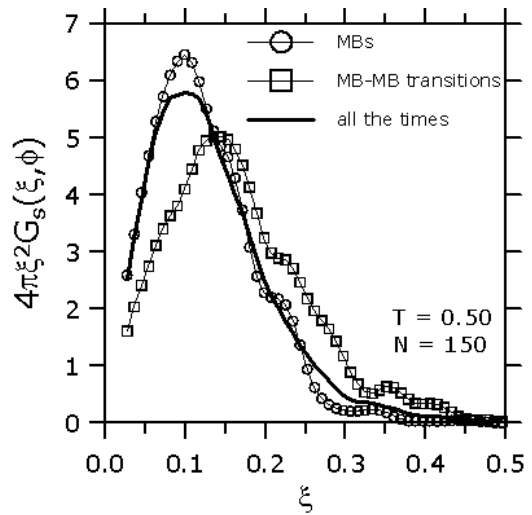}
\caption{Radial probability density $4 \pi \xi^2 G_{\rm s}(\xi,\phi)$ of finding a particle with a (root-squared) propensity change of magnitude $\xi$ in a $\phi=40$ time interval. Solid line: average over the whole run. Line with circles: average over time intervals within 4 selected MBs. Line with squares: average over 4 selected MB-MB transitions. Selected times are the same as in \cite{appignanesi_06}. Because of the noise in the data, these functions have been smoothed.}
\label{fig4}
\end{figure}

In order to better quantify the role of d-clusters and MB-MB transitions in changing the propensity map, we calculate the distribution of (root-squared) changes in the particle propensities, $\xi_i = |P_i(t+\phi)-P_i(t)|^{1/2}$. A large value of $\xi_i$ implies that particle $i$ has significantly changed its tendency to be mobile at the time $t+\phi$ with respect to its tendency at time $t$. The average of the distribution function of $\xi_i$ for all the 100 time intervals ($4 \pi \xi^2 G_{\rm s}(\xi,\phi)$), as depicted by the solid line in Fig.~\ref{fig4}, would represent an analogue of the self part of the van Hove function in the dynamics of the system (the probability that a particle after a time $\phi$ be found at distance $r$ from its original position). If we restrict this calculation to the times when the system is within an MB (and thus to the squared islands of Figs.~\ref{fig1} and \ref{fig3}), we get a similar curve (as shown in Fig.~\ref{fig4} in line with circles), since the system spends most of the time within MBs and thus this behavior dominates the whole distribution. However, if instead we calculate $\xi$ for the time intervals where an MB-MB transition occurs (the times when the system shows the crossings between islands in Figs.~\ref{fig1} and \ref{fig3}), the distribution (line with squares) is clearly displaced to the right, thus indicating that a great enhancement in the changes in particle propensity occurs at such times (as can also be learnt from Fig.~\ref{fig4}).

\begin{figure}[h!]
\includegraphics[width=1\linewidth]{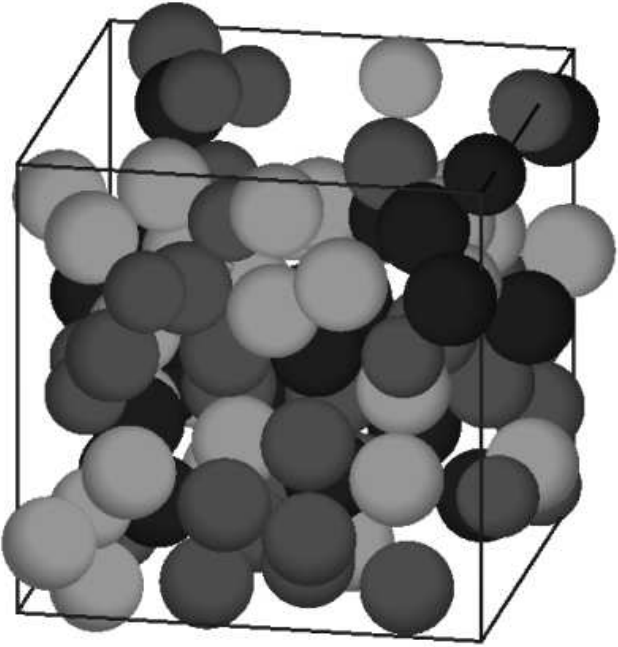}
\caption{Cluster of democratic particles (light- and dark-gray spheres) and high-propensity change (black and dark-gray spheres) for the MB-MB transition in the time interval [680,720] in the trajectory seen in Fig.~\ref{fig1}. Dark-gray particles are both democratic and have a high-propensity change.}
\label{fig5}
\end{figure}

We also calculated the fraction of particles that display large changes in $\xi$ and denoted this function as $m_\xi(t,\phi)$. As a threshold we use $\xi>0.21$ in order to have values comparable with the function $m(t,\phi)$ (that is $\int_{0.3}^\infty r^2 G_{\rm s}(r,\phi) {\rm d}r = \int_{0.21}^\infty \xi^2 G_{\rm s}(\xi,\phi) {\rm d}\xi = 0.16/(4\pi)$). In Fig.~\ref{fig2} we have included this function (light-gray-dashed line). Direct inspection of such plot shows the clear correlation that $m_\xi(t,\phi)$ exhibits with $\delta^2(t,\theta)$ and the function $m(t,\phi)$. The fraction of matches (function $M(t,\phi$), vertical bars) between particles that are both democratic and that exhibit a large propensity change (in a $\phi$ time interval) is also indicated.
Finally, to further show the good homology between democratic particles (which displacements are greater than 0.3 in a $\phi$ time interval) and particles that change significantly their propensity (particles with $\xi>0.21$ in a $\phi$ time interval) in a typical MB-MB transition, we show in Fig.~\ref{fig5} the spatial distribution of both kinds of particles for the MB-MB transition that occurs in the time interval [680,720] in the trajectory seen in Fig.~\ref{fig1}. Like the d-cluster consisting of democratic particles (light- and dark- gray spheres), the particles with high $\xi$ (black- and dark-gray spheres) are arranged in relatively compact clusters. Both clusters occupy the same region in space and many particles are both democratic and also have $\xi>0.21$ in the [680,720] time interval as can be seen in the figure (represented by dark-gray spheres).

\begin{figure}[h!]
\includegraphics[width=1\linewidth]{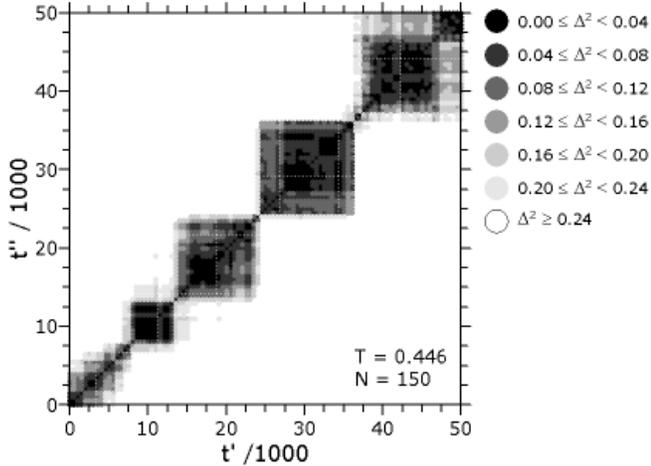}
\caption{Contour plot of the distance matrix $\Delta^2(t',t'')$ for 101 equally-spaced (500 time units) configurations of a trajectory at $T=0.446$. The gray levels correspond to the values that are given to the right.}
\label{fig6}
\end{figure}

\begin{figure}[h!]
\includegraphics[width=1\linewidth]{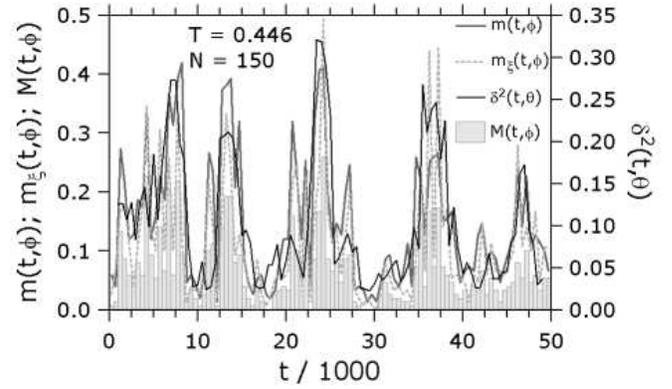}
\caption{
Solid-black line (right ordinate): Averaged-squared displacement $\delta^2(t,\theta)$ for the trajectory given in Fig.~\ref{fig6}.
Solid-dark-gray line (left ordinate): The function $m(t,\phi)$ which gives the fraction of democratic particles, i.e., particles with displacement $r > 0.3$ in the time interval $[t,t+\phi]$.
Dashed-light-gray line (left ordinate): The function $m_\xi(t,\phi)$ which gives the fraction of particles that changed their propensity more than $\xi = 0.21$ in the time interval $[t,t+\phi]$.
Vertical bars (left ordinate): The function $M(t,\phi)$ with gives the fraction of particles that are both democratic and have $\xi > 0.21$. The values of $\theta$ and $\phi$ are respectively 2000 and 500. Propensities are calculated for a $\phi=500$ time interval over ICs of 200 trajectories.}
\label{fig7}
\end{figure}

\begin{figure}[h!]
\includegraphics[width=1\linewidth]{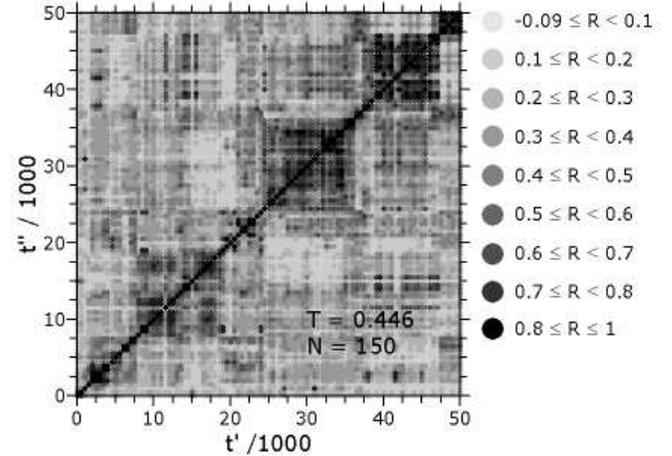}
\caption{Cross-correlation function $R(t',t'')$ between propensities calculated over ICs generated from configurations at times $t'$ and $t''$ that belong to the trajectory given in Fig.~\ref{fig6}. The propensities are calculated for a $\phi=500$ time interval. The gray levels correspond to the values that are given to the right.}
\label{fig8}
\end{figure}

As a final remark, we should like to mention that the behavior above expounded is characteristic of dynamical heterogeneities and thus emerges within the temperature interval where these heterogeneities are present. At temperatures lower than $T=0.5$ the behavior is very similar (in fact the results are a bit neater since the dynamical heterogeneities are more conspicuous) but the timescales grow since $t^*$ increases with decreasing temperature (400 for $T=0.50$ and 5000 for $T=0.446$ \cite{kob_95}). In Figs.~\ref{fig6}, \ref{fig7} and \ref{fig8} we show the situation for $T=0.446$.
We mention that the functions in Fig.~\ref{fig4} ($T=0.50$) are very similar to those for $T=0.446$, and the threshold $r=0.3$ for democratic particles in a $\phi=40$ time interval for $T=0.50$ (10 \% $t^*(T)$) is also reasonable in a $\phi=500$ time interval for $T=0.446$ (10 \% $t^*(T)$). Thus, we kept using the value $\xi>0.21$ for a criterium of high propensity change in a $\phi=500$ time interval for $T=0.446$.

On the other hand, at higher temperatures (higher than approximately $T=0.6$), the system looses the dynamical heterogeneities and becomes homogeneous and diffusive. Consequently, the distance matrix $\Delta^2$ lacks the island structure (as the non-gaussian parameter $\alpha_2$ begins to vanish) and the relaxation becomes diffusive, without the presence of d-clusters \cite{appignan_05}. Additionally, as showed in Ref. \onlinecite{widmer-cooper_04}, the propensities also get uniform at high temperature.

\section{Conclusions}

The isoconfigurational ensemble has provided us with a means of determining regions of the sample that present different tendency for mobility, thus enabling to measure the influence that the local structure exerts on the dynamics \cite{widmer-cooper_07,widmer-cooper_04}. The presence of relatively compact regions with high propensity for motion indicates domains made up of particles with high tendency to be mobile wherein the events responsible for the $\alpha$ relaxation occur \cite{PRL_2,ARF07,frechero_06}.
The fact that the spatial distribution of the very short time propensity \cite{ARF07,widmer-cooper_07} is sufficient to signal such regions and that some of such high propensity particles will take part of a d-cluster in the different IC trajectories at a later time \cite{ARF07,frechero_06} would mean that these particles are not at ease (not blocked) in the local structure and try for some time without success to relax their condition until eventually, a collective motion of many of them is able to perform a large scale relaxation event. While a rigorous complete comparison is not possible at this stage, this behavior seems to be in accord with an appealing very recently proposed soft mode explanation of glassy dynamics \cite{widmer-cooper_08}.
In such scenario \cite{widmer-cooper_08}, the high propensity regions present at a given configuration and the irreversible reorganization regions that occur later on are causally correlated with the localized low-frequency normal modes of such configuration and which persist for timescales on the order of $t^*$.
We note that the black islands of Figs.~\ref{fig3} and \ref{fig8} indicate that the spatial distribution of propensity (the tendency of the particles to move when evaluated at very short times) persist for times $|t'-t''|\sim t^*$. Within such timescales, the system changes many times the inherent structure (IS, local minimum or basin of attraction in the potential energy surface), an event that entails a small localized particle rearrangement, but has not been able to perform a collective long-range rearrangement characteristic of a d-cluster since it has been confined within the local metabasin, a collection of structurally very similar configurations. The d-cluster, which brings the system out of such MB, occurs within the high propensity region \cite{ARF07,frechero_06} and has been related to the soft modes present in the system at such times \cite{brito_07,coslovich_07, biroli_06}.
To summarize, in this work we have shown that not only the local structure poses its constraints on the dynamics of glassy systems, but that the opposite is also valid, that is, the dynamical events responsible for the $\alpha$ relaxation clearly modify the propensity pattern. Our results demonstrate that when a d-cluster (and the corresponding MB-MB transition) occurs, the spatial variation of propensity is reformulated (high- and low-propensity regions clearly de-correlate), thus making evident the mutual interplay between structure and dynamics. 

\begin{acknowledgments}
Financial support from ANPCyT, SeCyT and CONICET is gratefully acknowledged. G.A.A. is research fellow of CONICET. J.A.R-F and L.M.A thank CONICET for a fellowship.\\
\end{acknowledgments}


\begin{thebibliography}{10}
\bibitem{footnote1}
We use reduced units as in Ref.~\onlinecite{appignanesi_06}.
\end{thebibliography}
\end{document}